\documentclass[12pt]{iopart}
\usepackage[utf8]{inputenc}
\usepackage{microtype}
\usepackage{graphicx}
\expandafter\let\csname equation*\endcsname\relax
\expandafter\let\csname endequation*\endcsname\relax
\usepackage{amsmath}
\usepackage{amssymb}
\usepackage[colorlinks,linkcolor=blue,urlcolor=blue,citecolor=blue]{hyperref}
\usepackage{url}
\usepackage{soul}
\usepackage{mathptmx}
\usepackage{times}
\usepackage{epsfig,amsopn}
\usepackage{graphicx}
\usepackage{bm,amsmath,amssymb}
\usepackage[sort, numbers]{natbib}
\usepackage{braket}
\usepackage{float}
\usepackage{enumerate}
\usepackage{hyperref}
\usepackage[normalem]{ulem}
\allowdisplaybreaks[1]

\DeclareMathAlphabet{\mathcal}{OMS}{cmsy}{m}{n}

\begin{document}
\def\be{\begin{equation}}
\def\ee{\end{equation}}
\def\bea{\begin{eqnarray}}
\def\eea{\end{eqnarray}}
\def\f{\frac}
\def\l{\label}
\def\nn{\nonumber}

\definecolor{dgreen}{rgb}{0,0.7,0}
\def\redw#1{{\color{red} #1}}
\def\green#1{{\color{dgreen} #1}}
\def\blue#1{{\color{blue} #1}}
\def\brown#1{{\color{brown} #1}}

\newcommand{\aref}[1]{Appendix~\ref{#1}}%
\newcommand{\sgn}[1]{\mathrm{sgn}({#1})}%
\newcommand{\erfc}{\mathrm{erfc}}%
\newcommand{\Erf}{\mathrm{erf}}%
\newcommand{\rom}[1]{\uppercase\expandafter{\romannumeral #1\relax}}
            
\date{}

\title{\textit{Sokoban} percolation on the Bethe lattice}

\author{Ofek Lauber Bonomo, Itamar Shitrit and Shlomi Reuveni}

\address{School of Chemistry, Center for the Physics \& Chemistry of Living Systems, Ratner Institute for Single Molecule Chemistry, and the Sackler Center for Computational Molecular \& Materials Science, Tel Aviv University, 6997801, Tel Aviv, Israel}
\ead{ofekzvil@mail.tau.ac.il, itamarshtrit@mail.tau.ac.il and shlomire@tauex.tau.ac.il}

\date{\today}


\begin{abstract}
\noindent 
`With persistence, a drop of water hollows out the stone' goes the ancient Greek proverb. Yet, canonical percolation models do not account for interactions between a moving tracer and its environment. Recently, we have introduced the \textit{Sokoban} model, which differs from this convention by allowing a tracer to push single obstacles that block its path. To test how this newfound ability affects percolation, we hereby consider a Bethe lattice on which obstacles are scattered randomly and ask for the probability that the \textit{Sokoban} percolates through this lattice, i.e., escapes to infinity. We present an exact solution to this problem and determine the escape probability as a function of obstacle density. Similar to regular percolation, we show that the escape probability undergoes a second-order phase transition. We exactly determine the critical obstacle density at which this transition occurs and show that it is higher than that of a tracer without obstacle-pushing abilities. Our findings assert that pushing facilitates percolation on the Bethe lattice, as intuitively expected. This result, however, sharply contrasts with our previous findings on the 2D square lattice. There, the \textit{Sokoban} cannot escape --- not even at densities well below the percolation threshold. We discuss the reasons behind this striking difference, which calls for a deeper and better understanding of percolation in the presence of tracer-media interactions.

\end{abstract}

\maketitle

\section{Introduction}

Alfréd Rényi humorously remarked that a mathematician functions as a contraption designed to convert coffee into theorems \cite{schechter1998mathematical}. Though the precise ritual of brewing one's morning cup may vary, the essence lies in the process of extraction. This requires water to find its way through the ground coffee beans and into the cup. But can they do so effectively? Percolation theory emerged as an attempt to tackle this and similar questions mathematically. It can thus be ironically described as a concerted effort to convert coffee into theorems about coffee. The existing body of literature on percolation is vast {\cite{shante1971introduction,redner1982directed, sokolov1986dimensionalities, havlin1987diffusion, sahimi1994applications, ben2000diffusion, schwartz2002percolation, stauffer2018introduction} and it includes some notable recent developments \cite{derenyi2005clique,Buldyrev2010, gallos2012small,karrer2014percolation, radicchi2015percolation, li2015percolation,  sammartino2023percolation}.  

Broadbent \& Hammersley initiated modern percolation theory by introducing the bond percolation problem \cite{broadbent1957percolation}. They envisioned a three-dimensional cubic lattice within sites are interconnected by bonds. These bonds can be open to fluid flow with some probability. Alternatively, they can be closed, thus blocking the flow, with the complementary probability. Water will percolate through the lattice provided a spanning cluster of open bonds exists. Such a cluster is then referred to as the percolation cluster.

The concept of percolation is not limited to bonds and it is also customary to think of sites as the percolating elements. To illustrate this, consider a lattice where $\rho$ represents the probability of a site being blocked or obstructed. High values of $\rho$ are characterized by small islands of vacant sites, sparsely scattered on the lattice. Conversely, as $\rho$ diminishes, these islands expand until --- at sufficiently low $\rho$ --- a percolating cluster of contiguous vacant sites is formed. The transition between these two cases occurs at a critical density, known as the percolation threshold. For most lattices, however, exact determination of this critical density is not possible, and one often resorts to numerical methods instead \cite{ stauffer2018introduction, ben2000diffusion}.

The Bethe lattice is a notable exception where the percolation problem is exactly solvable. It consequently serves as a convenient analytical testbed to study percolation and related problems. Here, we take advantage of this proven workhorse to study a model of percolation with tracer media interactions. The motivation for our model comes from a simple observation: when pressurized water is pushed through the coffee basket of an espresso machine, it does not leave the original arrangements of coffee grains intact. Rather, water paves its way through the grains, often displacing them from their original positions. While the ability of water to push obstacles that obstruct its flow may be limited, it must have some effect on percolation. This idea is illustrated in Fig. \ref{Illustration} which compares percolation with and without obstacle pushing.

The importance of such interactions can also be understood by taking a random walks perspective on percolation. This was introduced by Pierre-Gilles de Gennes who coined the term “ant in a labyrinth” (AIL) to describe the motion of a random walker in disordered media \cite{de1976percolation} (similar ideas were explored by Brandt \cite{Brandt}, Kopelman \cite{diestler1976exciton}, and Mitescu \& Roussenq \cite{mitescu1976fourmi}). How the ant’s motion is affected by the density of surrounding obstacles is then of particular interest. Considering site percolation on a lattice, it is clear that when the obstacle density $\rho$ is small, most sites are unoccupied and the ant’s motion is almost unobstructed. However, when $\rho$ is large, most sites are occupied by obstacles, rendering the ant's motion extremely restricted. 

Assuming the ant cannot affect surrounding obstacles, e.g., push, pull, or otherwise change their position, the transition between restricted motion and free diffusion identifies with the critical obstacle density of site percolation \cite{ stauffer2018introduction, ben2000diffusion}. Interestingly, very close to the critical density, the ant exhibits anomalous, sub-diffusive behavior while roaming the percolating cluster which is a fractal \cite{havlin1987diffusion, ben2000diffusion, gefen1983anomalous, klafter2011first}.


\begin{figure*}[t!]
\centering
\includegraphics[width=0.75 \linewidth]{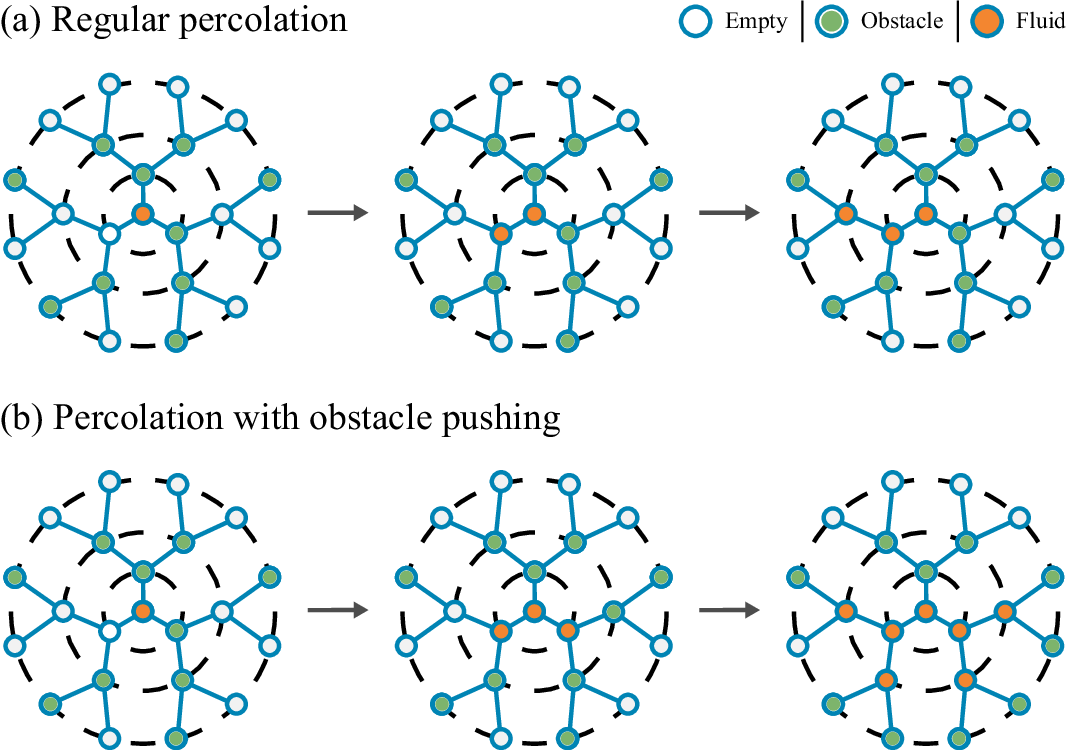}
\caption{Percolation with and without obstacle pushing. Clear differences are seen between the two cases. Panel (a) Regular percolation on the Bethe lattice. Obstacles are pinned and cannot move. Panel (b) Percolation where obstacles can be pushed by the force of the fluid. Illustrated here is a scenario where the flow is only strong enough to push single obstacles.}
\label{Illustration}
\end{figure*}


The assumption that the motion of a tracer does not affect the distribution of obstacles around it is prevalent, but not always valid.  Active tracers, e.g., animals, microorganisms, and biological `machines,' can plow their way through crowded environments by pushing obstacles that stand in their path. For example, light-activated Janus particles can imprint trails on their surroundings, thus forming a memory of past events  \cite{dias2023environmental}. Motorized robots, and self-propelled camphor boats, moving in an arena of inert—yet pushable—obstacles have been shown to affect their environment in a similar way \cite{biswas2020first,altshuler2023environmental}. It has also been shown that the formation of trails, and hence of environmental memory, can facilitate diffusion and shorten the first-passage time of a tracer to a target \cite{altshuler2023environmental}.

The experimental works described above reveal a series of non-trivial effects arising from tracer-media interactions. These hint that the percolating abilities of a “pushy” tracer could significantly deviate from those of the non-pushy ant. To find out, we have recently introduced a minimalist model aimed to capture the effect of obstacle pushing \cite{bonomo2023loss}. In this model, a random walker is given the ability to push single obstacles that block its path. Namely, the walker can move forward by pushing a single obstacle in its direction of motion, provided that the next site is unoccupied. Thus, in contrast to the ant, our walker modifies its surroundings as it moves through it. Consequently, the initial obstacle configuration is generally different than the one that emerges at later times.


\begin{figure*}[t!]
\centering
\includegraphics[width=0.75 \linewidth]{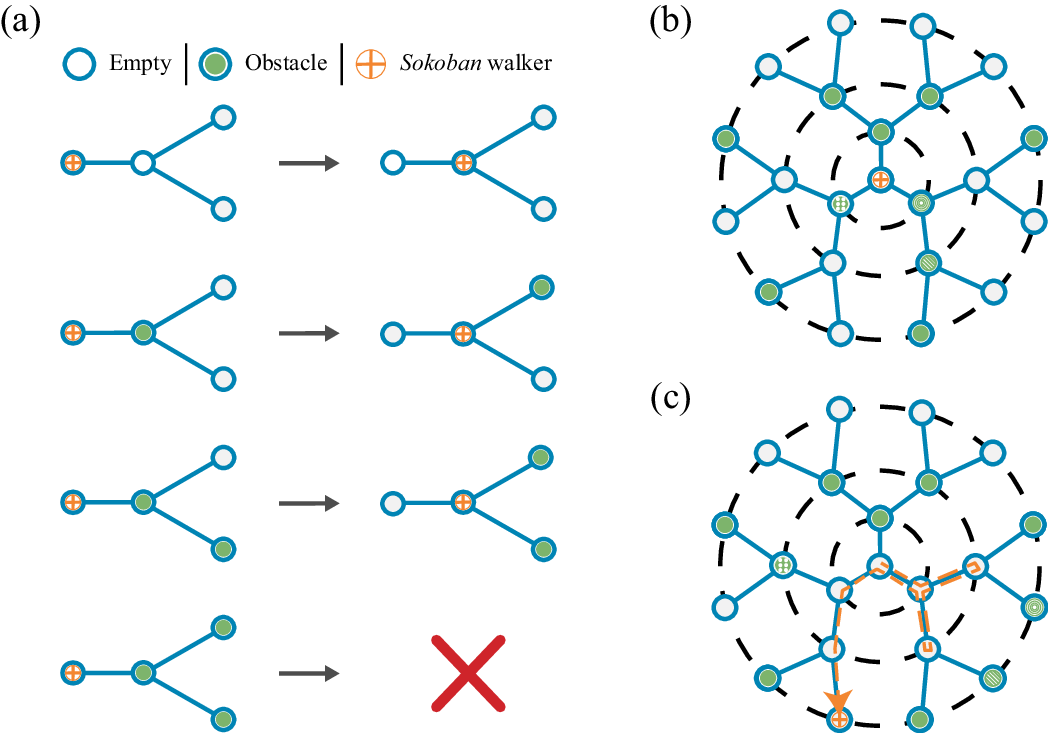}
\caption{The \textit{Sokoban} random walk on the Bethe lattice. Panel (a) Laws of motion. White circles indicate unoccupied nodes and green circles indicate nodes occupied by obstacles. The walker, marked by an orange cross, has two feasible moves: it can step into an unoccupied node, or step into an occupied node by pushing the obstacle that occupies it into an unoccupied node in the direction of motion. If there is more than one unoccupied node into which an obstacle can be pushed, the obstacle is pushed into one of the unoccupied nodes at random. The \textit{Sokoban} cannot push more than one obstacle at a time. Thus, an occupied node whose children nodes are all occupied forms an impassable roadblock. Panel (b) An example of an initial configuration of a Bethe lattice partially covered with obstacles. The \textit{Sokoban} is placed at the central node and obstacles are scattered, with probability $\rho$, at all other nodes. Obstacles that are pushed as the walker takes the trajectory illustrated in panel (c) are distinguished by different textures. Panel (c) A possible trajectory of the \textit{Sokoban} random walk. At each time step, the walker chooses between all feasible moves with equal probability, obeying the laws of motion in panel (a).}
\label{Rules}
\end{figure*}


We named our model for the pushy ant: the \textit{Sokoban} random walk. This was done in homage to the video game \textit{Sokoban} (Japanese for warehouse keeper), which was created in 1981 by Hiroyuki Imabayashi \cite{wikipediaSokoban}. The premise of the game is simple. Playing as the warehouse keeper, one pushes boxes around in a warehouse, with the goal of transporting them to marked storage locations. The rules of the game are similar to the laws of motion of the \textit{Sokoban} random walk. These are illustrated, on the Bethe lattice, with the aid of Fig. \ref{Rules}a. An illustration of the \textit{Sokoban} random walk is given in panels (b) and (c) of Fig. \ref{Rules}, where we present some initial obstacle configuration and give a sample trajectory, respectively. Note that in our example the \textit{Sokoban}  is not bound to the origin, even though all nodes around it are occupied by obstacles. The difference between the pushy \textit{Sokoban} and the non-pushy ant is further illustrated in Supplementary Video \#1.

The equivalence between regular percolation and the ant in a labyrinth model is well established; and similar equivalence exists between the pushy percolation illustrated in Fig. \ref{Illustration} and the \textit{Sokoban} random walk in Fig. \ref{Rules}. Thus, moving forward, we will adopt the random walks perspective to study the effect of pushing on percolation in the Bethe lattice. The remainder of this paper is structured as follows. In Sec. \ref{Preliminaries}, we review known results for the escape probability and critical density of an ant in a Bethe lattice labyrinth. In Sec. \ref{Model}, we proceed  to study  \textit{Sokoban} percolation on the Bethe lattice. First, we estimate the critical density using a mean-field argument. Next, we compute the escape probability and critical density of the \textit{Sokoban} exactly, and compare them to those of the non-pushy ant. We corroborate the results with the aid of an independent algorithm that can estimate the escape probability to arbitrary precision. This algorithm is developed in \ref{Numerics}. We end in Sec. \ref{Conclusions} with conclusions and outlook.


\section{Preliminaries: Ant in a Bethe lattice labyrinth}
\label{Preliminaries}

Before starting our analysis of the \textit{Sokoban} random walk, we recall essential facts about the Bethe lattice and known results for percolation on it. 

The Bethe lattice is an infinite tree where all nodes have the same number of neighbors $z$. This number is known as the coordination number or the degree of the lattice \cite{bethe1935statistical}. While all nodes in the Bethe lattice are identical by construction, it is convenient to define a central node (root). Other nodes on the lattice can then be grouped by their generation number, namely, their distance from the central node. The number of nodes at generation $g=1$ is given by $z$, i.e., the number of children nodes that are directly connected to the root, which is at $g=0$. Note that each node in generation $g\geq1$ has a single ‘parent’ node (a directly connected node at generation $g-1$). It also has $z-1$ ‘children’ nodes (directly connected nodes at generation $g+1$). Therefore, the number of nodes in generation $g+1$ is related to the number of nodes in generation $g \geq 1$ by $N_{g+1}=(z-1)N_{g}$. Iterating, we obtain $N_g=z(z-1)^{g-1}$ for the number of nodes in generation $g \geq 1$. Due to its distinctive topological features, and loop-less structure, models on the Bethe lattice are often easier to solve than on other lattices \cite{stauffer2018introduction, Baxter1982-qb}. A prime example of this, percolation, is now reviewed. 

Consider a Bethe lattice with coordination number $z$, where each node (except for the central node) has a probability $\rho$ to be occupied by an obstacle. In what follows we will also refer to $\rho$ as the obstacle density. A random walker is placed on the central node, which is unoccupied by convention. At each time step, the random walker takes a step to one of its \emph{unoccupied} neighboring nodes, if such exist, with equal probability. The walker cannot move into nodes already occupied by obstacles. Following Pierre-Gilles de Gennes  \cite{de1976percolation}, we refer to this random walk as an “ant in a Bethe lattice labyrinth”.

We are interested in the probability that the ant escapes from the central node to infinity. Namely, that it will \emph{eventually} reach at least one node in any generation $g=1,2,3,...$ of the Bethe lattice. Note that this probability, which we denote as $P_{\infty}^{AIL}$, is identical to the probability that the random walker starts on an infinite cluster of unoccupied nodes. Next, we show that there is a critical obstacle density, $\rho_c ^{AIL}$, below which $P_{\infty}^{AIL} > 0$. Also, above this critical density $P_{\infty}^{AIL} = 0$. We do so by following the footsteps of \cite{stauffer2018introduction}, generalizing the calculation given there from the particular case of $z = 3$ to a general coordination number $z$. 

We start by considering the central node of the Bethe lattice. For each one of its $z$ children at generation $g = 1$, we define the corresponding branch as the sub-graph consisting of: the central node, the specific child node chosen, all this child node's children, grandchildren, and so on to infinity. Consider a random walker that starts on the root of such a branch,  isolated from the rest of the lattice. We define the probability $E^{AIL}$ as the probability that the walker escapes to infinity through this branch. The probability of escaping to infinity on the original Bethe lattice follows immediately 
\bea 
P_{\infty} ^{AIL} = 1 - \left (Q^{AIL} \right )^z, \label{PercolationOPEQ}
\eea 
where $Q^{AIL} = 1 - E^{AIL}$. 

To find $Q^{AIL}$, we observe the corresponding branch as defined above and consider the two relevant scenarios: (i) the specific child node chosen is occupied; and (ii) the specific child node chosen is unoccupied, but the walker cannot escape to infinity through any of its children nodes. The probability $Q^{AIL}$ is then given by
\bea
Q^{AIL} = \rho + (1 - \rho)\left (Q^{AIL} \right )^{z-1}. \label{PercolationQEQ}
\eea 
For a general $z$ and a given $\rho$, Eq. \eref{PercolationQEQ} can be solved numerically to get the probability $Q^{AIL}$. Substituting back into \eref{PercolationOPEQ} gives the escape probability.  

Although  \eref{PercolationQEQ} cannot be solved analytically for a general value of $z$, the critical density $\rho_{c} ^{AIL}$ can be computed exactly. To do this, observe that near criticality $Q^{AIL} \rightarrow 1$. Thus, near the critical density, one can expand $\left (Q^{AIL} \right )^{z-1}$ on the right-hand side of \eref{PercolationQEQ} in the small parameter $1-Q^{AIL}$. Doing so, we obtain 
\begin{align}
\left (Q^{AIL} \right )^{z-1} &= [1 - (1 -Q^{AIL})]^{z - 1} \nn \\
&\simeq 1 - (1 - Q^{AIL}) (z - 1) + \frac{1}{2} (1 - Q^{AIL})^2 (z - 2) (z - 1),\label{QZApprox}
\end{align}
which is exact to second order. 
We then substitute the above approximation into \eref{PercolationQEQ} and solve for $Q^{AIL}$ to get the following non-trivial solution
\bea 
Q^{AIL} \simeq 1-\frac{2 [(1-\rho) (z-1)-1]}{(1-\rho) (z-2) (z-1)}. \label{QApprox} 
\eea 
The critical density can now be derived by substituting \eref{QApprox} into \eref{PercolationOPEQ} and setting $P_{\infty} ^{AIL}=0$. Solving for $\rho$, yields
\bea 
\rho_{c} ^{AIL} = 1 - \frac{1}{z-1}, \label{SimpleCrit} 
\eea
\noindent which is exact.

Noteably, the critical density in \eref{SimpleCrit} can also be obtained using a mean-field argument  \cite{stauffer2018introduction}. To do this, consider a random walker that has already made it to a node that resides at generation $g>0$. Having at least one of the $z-1$ children of this node unoccupied, guarantees that the random walker will be able to move one step further away from the central node, and this argument can be iterated. Mean field analysis replaces this \emph{exact} condition with the requirement that there is at least one unoccupied child node \emph{on average}. Since the mean number of unoccupied children nodes is given by $(1 - \rho) (z-1)$, we have $(1 - \rho_c) (z-1) = 1$ and solving for $\rho_c$ yields the critical density in \eref{SimpleCrit}. 

Before concluding this section, we note that the behavior of the escape probability near criticality can also be determined. To do so, we first note that near the critical density, we can approximate \eref{PercolationOPEQ} as $P_{\infty} ^{AIL} = 1 - \left (Q^{AIL} \right )^z \simeq (1 - Q^{AIL})z$. Substituting  \eref{QApprox} into this relation and expanding to first order in the small parameter $\rho_{c} ^{AIL}-\rho$, yields 
\bea
P_{\infty} ^{AIL} \simeq \frac{2z(z - 1)}{z-2}(\rho_{c} ^{AIL} - \rho). \label{PercolationOPScaling}
\eea 
In Sec. \ref{Exact Results}, we will compare the results in \eref{SimpleCrit} and \eref{PercolationOPScaling} with analogous results obtained for the \textit{Sokoban} random walk on the Bethe lattice. 

\section{\textit{Sokoban} random walk on the Bethe lattice}
\label{Model}
We now proceed to explore how the ability to push obstacles affects percolation on the Bethe lattice. We consider a Bethe lattice with an arbitrary coordination number $z$ and obstacle density $\rho$. We place a \textit{Sokoban} random walker at the central node of this lattice and allow it to take steps according to the laws of motion that were illustrated in Fig. \ref{Rules}a. Similar to Sec. \ref{Preliminaries}, we are interested in calculating $P_{\infty} ^{\textit{Sokoban}}$. Namely, the probability that the \textit{Sokoban} random walk escapes from the central node to infinity. We are also interested in finding the critical obstacle density $\rho_{c} ^{\textit{Sokoban}}$ above which $P_{\infty} ^{\textit{Sokoban}}$ vanishes. Before providing an exact solution to this problem, which we will obtain by adapting the approach in Sec. \ref{Preliminaries} to the \textit{Sokoban} random walk, we will present a mean-field argument for the calculation of the critical density.  

\subsection{Mean field analysis}
\label{MeanField}
Consider a \textit{Sokoban} random walker that has already made it to a node that resides at generation $g>0$. According to the laws of motion given in Fig. \ref{Rules}a, the walker can only push one obstacle at a time. Thus, having at least one unoccupied \emph{grandchild} node, out of the $(z-1)^2$, will guarantee that the \textit{Sokoban} can move to generation $g+1$, i.e, one step further away from the central node. Mean field analysis replaces this \emph{exact} condition with the requirement that there is at least one unoccupied grandchild node \emph{on average}.  Since the mean number of unoccupied grandchildren nodes is given by $(1 - \rho)(z - 1)^2$, we have $(1 - \rho_c)(z - 1)^2 = 1$ and solving  yields  
\bea 
\rho_{c} ^{\textit{Sokoban}} = 1 - \frac{1}{(z-1)^2},\label{SokoCrit} 
\eea
which despite the approximation turns out to be exact as we show below. Note that the critical density for the \textit{Sokoban} is higher than the critical density in \eref{SimpleCrit}. This is in agreement with the intuition that the ability to push obstacles enables the \textit{Sokoban} to venture further than the AIL. 

\subsection{The escape probability and critical density}
\label{Exact Results}

In this section, we calculate the probability that the \textit{Sokoban} random walk
escapes from the central node to infinity. To do so, we generalize the calculation in Sec. \ref{Preliminaries} such that it would take into account the \textit{Sokoban}'s ability to push obstacles.

Consider a \textit{Sokoban} random walker that starts its motion from the central node of a Bethe lattice with obstacle density $\rho$. Similarly to the calculation done for the ant in a labyrinth, we would like to start by writing an equation for the escape probability $P_{\infty}^{\textit{Sokoban}}$. For the ant, this probability was expressed in terms of the probability $Q^{AIL} = 1 - E^{AIL}$ defined in Sec. \ref{Preliminaries}. Recall, that $E^{AIL}$ is the probability that the ant escapes to infinity through a given branch. The fact that the ant cannot push obstacles, simplifies the calculation of this probability as it is impossible to escape through a branch that originates from an occupied node. In contrast, the \textit{Sokoban} plows its way through obstacles and may escape even from a branch that originates in an occupied node. Thus, to calculate $E^{\textit{Sokoban}}$, one needs to consider two different possibilities: (i) escape to infinity through a branch originating from an \emph{unoccupied} node; and (ii) escape to infinity through a branch originating from an \emph{occupied} node. 

To take both scenarios into account, we define two auxiliary probabilities (also see Fig. \ref{ProbIllusFig}): (i) $P_{Empty}$, the probability that the \textit{Sokoban} random walker escapes to infinity through a branch originating from an \emph{unoccupied} node; and (ii) $P_{Full}$, the probability that the \textit{Sokoban} random walker escapes to infinity through a branch originating from an \emph{occupied} node. The probability $Q^{\textit{Sokoban}}$ can then be written in terms of these auxiliary probabilities. This relation reads 
\bea
Q^{\textit{Sokoban}} = 1 - ((1 - \rho)P_{Empty} + \rho P_{Full}). \label{Q SOKOBAN}
\eea 
The escape probability of the \textit{Sokoban} from the Bethe lattice can then be written as
\bea 
\label{StrengthEq}
P_{\infty}^{\textit{Sokoban}} &=& 1- \left ( Q^{\textit{Sokoban}} \right )^z \nn \\ &=& 1 - \left(1 - ((1 - \rho)P_{Empty} + \rho P_{Full})\right)^z,
\eea 
which is the \textit{Sokoban} analog of Eq. \eref{PercolationOPEQ}.

We now turn to find $P_{Empty}$ and $P_{Full}$ which are still unknown. These probabilities obey 
\begin{equation}
\label{System}
\left\{
\begin{alignedat}{2}
  P_{Empty} \; = \; &1-\left(1-((1-\rho) P_{Empty}+\rho P_{Full}) \right)^{z-1}, \\
  P_{Full} \; = \; &1 - \rho^{z - 1} \\
  & -\sum _{n=0}^{z-2} \binom{z-1}{n} \rho^{n} (1-\rho)^{z-n-1}
   (1-P_{Empty})^{z-n-2} (1-P_{Full})^{n+1}.
\end{alignedat}
\right.
\end{equation}
The first equation in the system above can be understood similarly to \eref{StrengthEq} albeit a small difference. Here the second term is raised to the power of $z-1$, instead of $z$, since we consider branches that originate from nodes other than the central node. These have only $z - 1$ children. The second equation is derived by considering all the scenarios where the \textit{Sokoban} random walker does not escape to infinity through a branch originating from an occupied node. The first scenario, is captured by the $\rho^{z-1}$ term that accounts for the case where all the children of the considered node are also occupied. In this case, the \textit{Sokoban} cannot take a step into the considered node in question since the obstacle occupying it cannot be pushed forward. All other scenarios are given by the third term on the right-hand side of the equation. The sum accounts for the probability of escaping to infinity through any possible configuration that is not fully occupied. Each term in the sum is a product of two probabilities: (i) the probability of initially having $n<z-1$ occupied nodes out of the $z-1$ children nodes, and (ii) the probability of escaping through such a configuration after pushing the obstacle at the considered  node one generation up. 


\begin{figure}[t!]
\centering
\includegraphics[width=0.4 \linewidth]{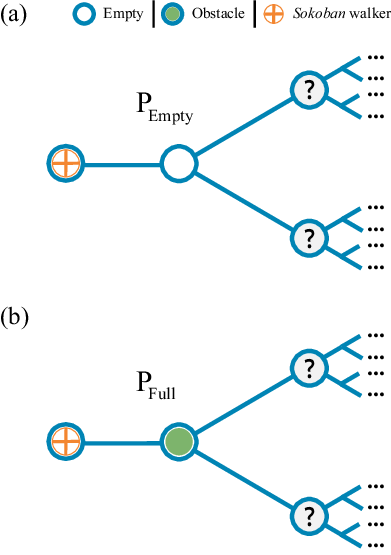}
\caption{Illustration of the auxiliary probabilities $P_{Empty}$ and $P_{Full}$. Panel (a): The probability $P_{Empty}$ is the probability that the \textit{Sokoban} (marked by an orange cross) escapes to infinity through a branch originating from an unoccupied node (marked by a white circle). Panel (b): The probability $P_{Full}$ is the probability that the walker escapes to infinity through a branch originating from an occupied node (marked by a green circle).}
\label{ProbIllusFig}
\end{figure}


Calculating the sum on the right-hand side of the second equation in \eref{System} using the binomial theorem yields
\begin{equation}
\label{SystemFin}
\left\{
\begin{alignedat}{2}
  P_{Empty} \; = \; &1-\left(1-((1-\rho) P_{Empty}+\rho P_{Full}) \right)^{z-1}, \\
  P_{Full} \; = \; &1 - \rho^{z - 1} \\
  & + \frac{\rho^{z-1} (1-P_{Full})^z-(1-P_{Full}) (1-(1-\rho) P_{Empty}-\rho P_{Full})^{z-1}}{1-P_{Empty}}.
\end{alignedat}
\right.
\end{equation}
Using the first equation in \eref{SystemFin}, we substitute $1-P_{Empty}$ for $\left(1-((1-\rho) P_{Empty}+\rho P_{Full}) \right)^{z-1}$ in the  \emph{numerator} of the second equation. Solving for $P_{Empty}$ yields 
\bea
\label{Prelation}
P_{Empty}=1-(1- P_{Full})^z.
\eea 

\noindent Replacing  $P_{Empty}$ in both sides of the first equation in \eref{SystemFin} with the expression in  \eref{Prelation} and solving for $\rho$ yields
\bea
\label{rhoSol}
1-\rho=\frac{1-(1-P_{Full})^{\frac{1}{z-1}}}{1-(1-P_{Full})^{z-1}}.
\eea 
Note that one can use the above equation to calculate $P_{Full}$ as a function of the obstacle density $\rho$ in a numerically exact manner. In addition, observe that the critical density of the \textit{Sokoban} random walk follows directly by taking the limit $P_{Full} \to 0$ on the right-hand side of \eref{rhoSol}. Doing so, e.g. via the L'Hôpital's rule or by expanding the numerator and denominator to first order, yields the critical density that was found in \eref{SokoCrit} using a mean-field argument. This shows that \eref{SokoCrit} is in fact exact.

\begin{figure*}[t!]
\includegraphics[width=1\textwidth]{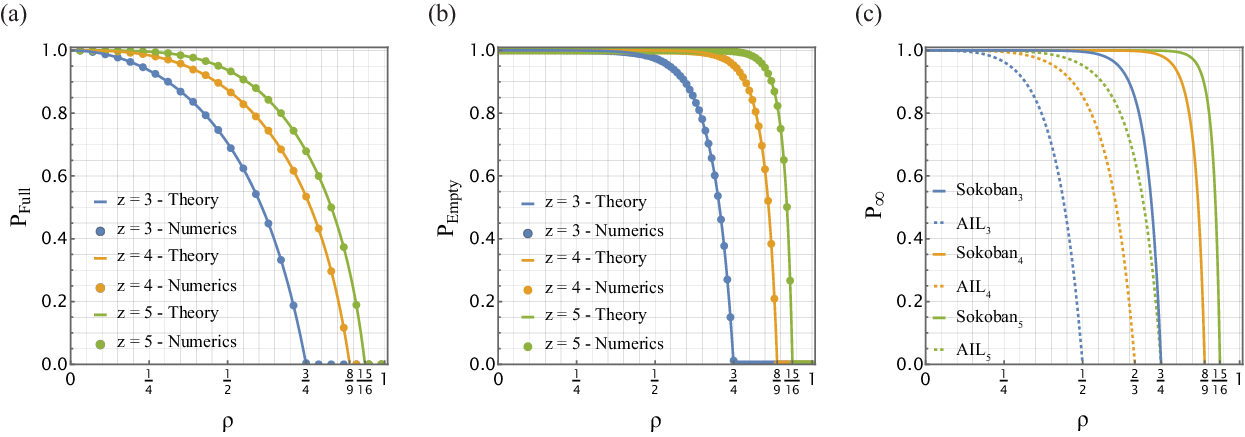}
\centering
\caption{Escape probabilities of the  \textit{Sokoban} random walk on the  Bethe lattice. Panels (a) and (b): $P_{Full}$ and $P_{Empty}$ respectively. These escape probabilities are plotted in solid lines vs. the obstacle density $\rho$, for lattices with coordination numbers $z=3,4,5$. Plots are made by numerically solving Eqs. (\ref{Prelation}) and (\ref{rhoSol}). Corroborating the theoretical results are circles that come from an independent algorithmic estimation of the escape probabilities (see \ref{Numerics} for details). Note that both escape probabilities vanish for obstacle densities that are above the critical density in (\ref{SokoCrit}). Panel (c): Comparison between the escape probability of the \textit{Sokoban} (solid lines) and the ant in a Bethe lattice labyrinth (dashed lines). Plots are made, using Eqs. \eref{StrengthEq} and \eref{PercolationOPEQ} respectively, as a function of the obstacle density $\rho$ and for lattices with coordination numbers $z=3,4,5$. Observe that   $P_{\infty}^{\textit{Sokoban}}>P_{\infty}^{\textit{AIL}}$ and that the critical density of the \textit{Sokoban} random walk is higher than that of the AIL.}
\label{ProbTheoryFig}
\end{figure*}

In Fig. \ref{ProbTheoryFig}a, we consider lattices of coordination numbers $z = 3,4,5$ and plot the theoretical prediction of \eref{rhoSol} for $P_{Full}$  vs. the obstacle density $\rho$ (solid lines). We further corroborate this result using values obtained from an independent algorithmic estimation of $P_{Full}$ (circles, see \ref{Numerics} for details). We observe a perfect match. Once at hand, $P_{Full}$ can be substituted into \eref{Prelation} to find $P_{Empty}$. In Fig. \ref{ProbTheoryFig}b, we plot the theoretical prediction of  \eref{Prelation} for $P_{Empty}$ (solid lines) vs. $\rho$. Once again, these results are corroborated successfully using values obtained from an independent algorithmic estimation of $P_{Empty}$ (circles, see \ref{Numerics} for details). 

Finally, substituting $P_{Full}$ and $P_{Empty}$ into \eref{StrengthEq} yields the escape probability. In Fig. \ref{ProbTheoryFig}c, we plot $P_{\infty}^{\textit{Sokoban}}$ vs. $\rho$ (solid lines), and compare to the escape probability $P_{\infty}^{\textit{AIL}}$ of an ant in a labyrinth (dotted lines). The latter is obtained from numerical solution of Eqs. (\ref{PercolationOPEQ}) and (\ref{PercolationQEQ}). It can be seen that $P_{\infty}^{\textit{Sokoban}}>P_{\infty}^{\textit{AIL}}$ in all cases. Namely, on the Bethe lattice, the pushing ability of the \textit{Sokoban} increases its probability to escape. Moreover, since the critical density of the \textit{Sokoban} is higher than that of the ant, there exists a range of densities for which the \textit{Sokoban} can escape the lattice with some probability while the ant cannot. For example, for $z=3$, this range is given by $1/2<\rho<{3/4}$. Note, throughout most of this range, the \textit{Sokoban} has a very high probability to escape, highlighting the significant impact of obstacle pushing.

\subsection{Behaviour near criticality}

We now study the behavior of the escape probabilities near the critical obstacle density. To do so, we go back to \eref{rhoSol} and expand its  right-hand side to first order in  $P_{Full}$, which is small near the critical density. This gives
\bea
\frac{1-(1-P_{Full})^{\frac{1}{z-1}}}{1-(1-P_{Full})^{z-1}} &\simeq& \frac{1}{(z-1)^2} + \frac{z(z-2)}{2 (z-1)^3} P_{Full} \nn \\
&=& 1 - \rho_{c}^{\textit{Sokoban}} + \frac{z(z-2)}{2 (z-1)^3} P_{Full}.
\eea
Substituting the above expansion in \eref{rhoSol} and solving for $P_{Full}$ yields
\bea
P_{Full} \simeq \frac{2 (z-1)^3}{z(z-2)} (\rho_{c}^{\textit{Sokoban}} - \rho), \label{Pfull scaling}
\eea 
near the critical density. Expanding $P_{Empty}$ in \eref{Prelation} to first order in  $P_{Full}$, and substituting the above relation we get
\bea
P_{Empty} \simeq \frac{2 (z-1)^3}{z-2} (\rho_{c}^{\textit{Sokoban}} - \rho). \label{Pempty scaling}
\eea 
From here we see that near the critical density $P_{Empty}$ changes with a slope that is $z$ times steeper than that of $P_{Full}$ (see panels (a) and (b) of Fig. \ref{ProbTheoryFig}).

We continue by substituting Eqs. (\ref{Pfull scaling}) and (\ref{Pempty scaling}) into \eref{StrengthEq}. Expanding to first order in $\rho_{c}^{\textit{Sokoban}} - \rho$ gives the following linear relation for the escape probability of the \textit{Sokoban} near the critical density
\bea
\label{SOKOBANScaling}
P_{\infty}^{\textit{Sokoban}} \simeq \frac{2z(z-1)^2}{z-2}(\rho_{c}^{\textit{Sokoban}} - \rho).
\eea 
It is interesting to compare the above relation to \eref{PercolationOPScaling}, which gives the corresponding relation for the escape probability of an ant in a Bethe lattice labyrinth. This comparison reveals that pushing does not change the critical exponent $P_{\infty} \sim (\rho_{c} - \rho)^{\beta}$, which is given by $\beta=1$ in both cases \cite{stauffer2018introduction}. However, near the critical density, the escape probability of the \textit{Sokoban} changes with a slope that is $z-1$ times steeper than that of the ant. This result is in agreement with   Fig. \ref{ProbTheoryFig}c  that shows extremely sharp transitions for the \textit{Sokoban} near the critical density. In Fig. \ref{ScalingLawFig}, we further corroborate this asymptotics by plotting  \eref{SOKOBANScaling} (dotted lines) alongside the exact escape probability from \eref{StrengthEq} (solid lines). It can be seen that the predicted asymptotics capture the behavior near the critical density.


\begin{figure}[t!]
\centering
\includegraphics[width=0.5 \linewidth]{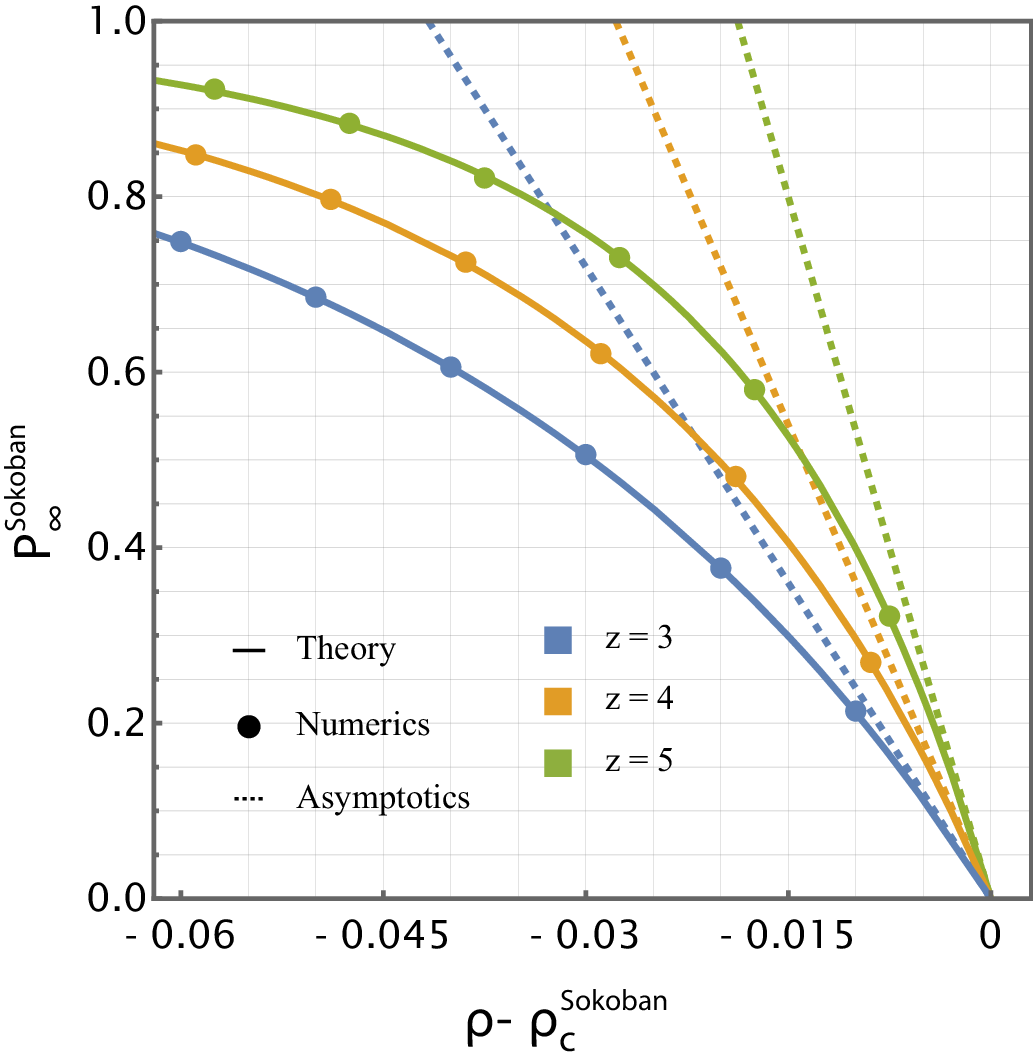}
\caption{The escape probability of the \textit{Sokoban} random walk near the critical density. Plots are made as a function of the deviation from the critical density, for lattices with coordination numbers $z=3,4,5$. Solid lines present exact results coming from \eref{StrengthEq} and dashed lines present the asymptotics of \eref{SOKOBANScaling}. The circles, which come from an independent algorithmic estimation of the escape probability (see \ref{Numerics} for details), further corroborate these results.}
\label{ScalingLawFig}
\end{figure}


\section{Summary and Outlook}
\label{Conclusions}

In this paper, we utilized the \textit{Sokoban} model to study the effect of interactions on percolation in disordered media.  We considered a randomly moving tracer --- the \textit{Sokoban} --- which can push single obstacles obstructing its path. We studied the motion of the \textit{Sokoban} on the Bethe lattice and determined exactly how its escape probability depends on obstacle density.

Our analysis shows that the ability to push obstacles significantly aids escape. Specifically, the \textit{Sokoban} consistently achieves a higher escape probability compared to the proverbial ant, which navigates the Bethe lattice without altering its environment. Both the \textit{Sokoban} and the ant exhibit a second-order phase transition of their escape probability. However, the \textit{Sokoban}'s transition occurs at a higher obstacle density than the ant's. As a result, for obstacle densities above the percolation threshold: the ant is trapped but the \textit{Sokoban} may still plow its way out to freedom.

Our findings support the intuition that pushing allows the \textit{Sokoban} to access places inaccessible to the non-pushy ant. However, this advantage disappears when the Bethe lattice is replaced with a 2D square lattice \cite{bonomo2023loss}. There, the \textit{Sokoban} cannot escape even when obstacle densities are well below the percolation threshold of the ant. This surprising behavior highlights the non-trivial effect that obstacle-pushing has on percolation and transport. It also warrants some further discussion.

We have recently conjectured that the \textit{Sokoban} random walk gets caged on the 2D square lattice even at arbitrarily low obstacle densities \cite{bonomo2023loss}. As it moves, the \textit{Sokoban} shovels obstacles from the area it visits to the periphery. With time, both the visited area and its periphery expand, but the periphery grows much more slowly. Consequently, all open spaces along the periphery are eventually filled with obstacles. In fact, we have shown that a double layer of peripheral obstacles is formed. This acts as a cage for the \textit{Sokoban} which cannot push more than one obstacle at a time.

The above argument was used to successfully explain quantitative relations that govern the caging of the \textit{Sokoban} random walk on the 2D square lattice \cite{bonomo2023loss}. On the Bethe lattice, the \textit{Sokoban} also displaces obstacles from visited sites to the periphery. Yet, on this lattice the number of periphery sites grows proportionally to the number of visited sites: for every newly visited site, there are $z-2$ sites added to the periphery.  Consequently, one can no longer guarantee that caging always occurs and the exact escape probability is given by Eq. \eref{StrengthEq}. 

Summarizing, we see that the ability to push obstacles affects percolation in a topology-dependent manner. On the loop-less Bethe lattice, it facilitates escape while on the 2D square lattice it leads to caging. How pushing affects percolation on other topologies, e.g., regular lattices in higher dimensions, fractal and random networks  \cite{havlin1987diffusion, ben2000diffusion, klafter2011first, newman2018networks, alma990029615420204146}, is still unknown. Addressing this knowledge gap should elucidate the relation between graph properties and the percolation of pushy tracers.  

\ack
This project has received funding from the European Research Council (ERC) under the European Union’s Horizon 2020 research and innovation program (Grant agreement No. 947731). 

\appendix
\section{Numerics}
\label{Numerics}

It is virtually impossible to estimate the escape probabilities of the \textit{Sokoban} using brute-force Monte Carlo simulations on the Bethe lattice. In this appendix, we develop instead an algorithm that exploits the self-similar structure of the lattice to provide numerical estimation of the escape probabilities $P_{Empty}$ and $P_{Full}$. Results coming from this algorithm are independent of the theory developed in the main text. They were hence used to corroborate the analytical predictions of Eqs. (\ref{Prelation}) and (\ref{rhoSol}).

\subsection{Estimating $P_{Empty}$}
\label{PemptyCalc}
Consider a branch of a Bethe lattice with coordination number $z$. Nodes on the lattice are occupied with probability $\rho$, except for the two nodes in generations $g=0$ and $g=1$ which are left empty. We place the \textit{Sokoban} walker in generation $g=0$, as illustrated in panel (a) of Fig. \ref{NumericsIllustration}.

To numerically estimate $P_{Empty}$, we define the following sequence of auxiliary probabilities for the system in panel (a) of Fig. \ref{NumericsIllustration}: $E_1 ^{Empty}, E_2 ^{Empty}, E_3^{Empty},...$. Here, $E_n ^{Empty}$ is the probability for a walker that reached a specific node in the $n$-th generation to escape through it to infinity, given it never returns to generation $n-1$. For convenience, we denote the complimentary probabilities $Q_n ^{Empty} = 1 - E_n ^{Empty}$. Note that $Q_1 ^{Empty} = 1 - P_{Empty}$, follows immediately. Thus, by estimating $Q_1 ^{Empty}$, one also gets $P_{Empty}$.


\begin{figure}[t!]
\centering
\includegraphics[width=0.4 \linewidth]{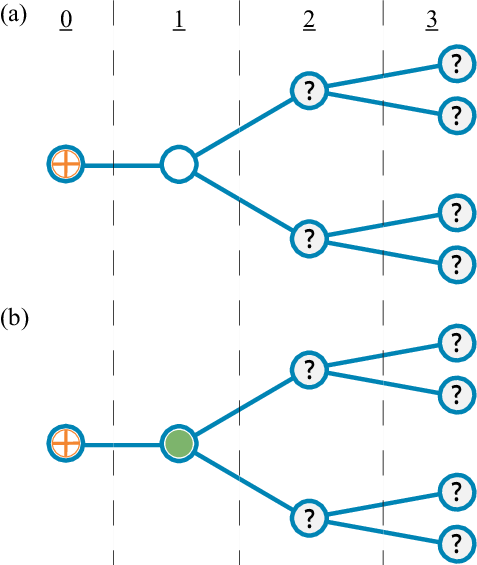}
\caption{Illustration of the systems used to calculate the auxiliary probabilities $Q_1 ^{Empty}$ and $Q_1 ^{Full}$. Generation numbers are indicated by overhead digits. Panel (a): A four generations Cayley tree branch (a finite-generation Bethe lattice) used for the calculation of the probability $Q_1 ^{Empty}$. By construction, the nodes in generations $0$ and $1$ are unoccupied, nodes in other generations are occupied with probability $\rho$. Panel (b): A four-generation Cayley tree branch used for the calculation of the probability $Q_1 ^{Full}$. By construction, the node in generation $0$ is unoccupied, the node in generation $1$ is occupied (marked by a green circle), and nodes in other generations are occupied with probability $\rho$.}
\label{NumericsIllustration}
\end{figure}


We start by calculating $Q_1 ^{Empty}$. This probability can be decomposed into two contributions: i) None of the children nodes of the node in the first generation are accessible to the walker. Namely, all of its children and grandchildren nodes are occupied; ii) The node in the first generation has $1 \leq m \leq z - 1$ accessible children nodes, through none of which the walker escapes to infinity. Thus, the probability $Q_1 ^{Empty}$ is given by the following formula
\bea
\label{Q(1)}
Q_{1} ^{Empty} &= \underbrace{R_{1} ^{Empty}(0)}_{a_1\left(R_{1} ^{Empty}(\cdot) \right)}+\underbrace{\sum_{m_1=1}^{z-1}R_{1} ^{Empty}(m_1)\left (Q_{2} ^{Empty} \right )^{m_1}}_{\epsilon_2 \left ( R_{1} ^{Empty}(\cdot), Q_2 ^{Empty} \right)},
\eea
where $R_{1} ^{Empty} (m)$, $0 \leq m \leq z - 1$, is the probability that exactly $m$ children nodes are accessible to the walker \emph{after} it arrives at the node in the first generation. Exploiting the recursive structure of the Bethe lattice, a similar formula relating the probabilities $Q_n ^{Empty}$ and $Q_{n + 1} ^{Empty}$, for $n \geq 1$, can be derived  
\bea
\label{Q(n)}
Q_n ^{Empty}&=&R_{n} ^{Empty}(0)+\sum_{m_n=1}^{z-1}R_{n} ^{Empty}(m_n)\left ( Q_{n + 1} ^{Empty} \right )^{m_n} \nn \\
&=& \sum_{m_n=0}^{z-1}R_{n} ^{Empty}(m_n)\left ( Q_{n + 1} ^{Empty} \right )^{m_n},
\eea 
where $R_{n} ^{Empty} (m)$, $0 \leq m \leq z - 1$, is the probability that exactly $m$ children nodes are accessible to the walker, \emph{after} it arrives at a specific node in the $n$-th generation. 

Setting $n=2$ in \eref{Q(n)} and substituting $Q_2 ^{Empty}$ into \eref{Q(1)}, the term $\epsilon_2$ can be rewritten as follows \footnotesize
\bea
\label{e2}
\epsilon_2 &=& \sum_{m_1=1}^{z-1}R_{1} ^{Empty}(m_1)\left (R_{2} ^{Empty}(0)+\sum_{m_2=1}^{z-1}R_{2} ^{Empty}(m_2)\left (Q_{3} ^{Empty} \right )^{m_2} \right )^{m_1} \nn \\
&=& \underbrace{\sum_{m_1=1}^{z-1}R_{1} ^{Empty}(m_1) \left (R_{2} ^{Empty}(0) \right) ^{m_1}}_{a_2\left(R_{1} ^{Empty}(\cdot), R_{2} ^{Empty}(\cdot) \right)} \nn \\
&&+ \underbrace{\sum_{m_1=1}^{z-1}R_{1} ^{Empty}(m_1) \sum_{k=1}^{m_1} \binom{m_1}{k} \left ( R_{2} ^{Empty}(0) \right) ^{m_1 -k} \left [ \sum_{m_2=1}^{z-1}R_{2} ^{Empty}(m_2)\left (Q_{3} ^{Empty} \right )^{m_2} \right ] ^{k}}_{\epsilon_3 \left ( R_{1} ^{Empty}(\cdot), R_{2} ^{Empty}(\cdot), Q_3 ^{Empty} \right)}.\nn \\ 
\eea 
\normalsize Substituting \eref{e2} into \eref{Q(1)} yields
\bea 
Q_1 ^{Empty} &=& a_1\left(R_{1} ^{Empty}(\cdot) \right) + a_2\left(R_{1} ^{Empty}(\cdot), R_{2} ^{Empty}(\cdot) \right) \nn \\
&& + \epsilon_3 \left ( R_{1} ^{Empty}(\cdot), R_{2} ^{Empty}(\cdot), Q_3 ^{Empty} \right).
\eea 

\noindent Repeating this procedure $g-2$ times yields the following formula for $Q_1 ^{Empty}$
\bea
\label{QsumA}
Q_1 ^{Empty} &=& \sum_{n=1}^{g - 1} a_n \left(R_{1} ^{Empty}(\cdot), ..., R_{n} ^{Empty}(\cdot) \right) \nn \\
&&+ \epsilon _{g} \left ( R_{1} ^{Empty}(\cdot), ..., R_{g - 1} ^{Empty}(\cdot), Q_{g} ^{Empty} \right).
\eea
Taking $g \rightarrow \infty$ in the above formula, we have
\bea
\label{QsumAInf}
Q_1 ^{Empty} = \sum_{n=1}^{\infty} a_n \left(R_{1} ^{Empty}(\cdot), ..., R_{n} ^{Empty}(\cdot) \right).
\eea
\noindent Note that this formula is \emph{independent} of the auxiliary probabilities $Q_1 ^{Empty}, Q_2 ^{Empty}, ...$. In practice, the infinite sum in \eref{QsumAInf} can be approximate by $Q_1 ^{Empty} \approx \sum_{n=1}^{g - 1} a_n$, with $g$ taken to be sufficiently large.

To calculate the aforementioned sum, we write \eref{QsumA} explicitly. Substituting \eref{Q(n)} repeatedly into \eref{Q(1)}, for increasing values of $n$, yields \small 
\bea 
\label{IterativeQ}
Q_1 ^{Empty} = \sum_{m_{1}=0}^{z-1}R_{1}^{Empty}(m_{1})\left(\cdots\left(\sum_{m_{g-1}=0}^{z-1}R_{g-1}^{Empty}(m_{g-1})\left(Q_{g}^{Empty}\right)^{m_{g-1}}\right)^{m_{g-2}}\cdots\right)^{m_{1}} . \nn \\
\eea 
\normalsize Equation (\ref{QsumA}) asserts that the above formula for $Q_1 ^{Empty}$ can be decomposed into two contributions, $\sum_{n=1}^{g - 1} a_n$ and $\epsilon_g$, where the former provides an approximation for $Q_1 ^{Empty}$ in which $Q_g ^{Empty}$ does not appear. The following formula follows immediately \small
\bea
\label{suma}
Q_1 ^{Empty} &\approx& \sum_{n=1}^{g - 1} a_n = \nn \\
&& \sum_{m_{1}=0}^{z-1}R_{1}^{Empty}(m_{1})\left(\cdots\left(\sum_{m_{g-2}=0}^{z-1}R_{g-2}^{Empty}(m_{g-2})\left(R_{g-1}^{Empty}(0)\right)^{m_{g-2}}\right)^{m_{g-3}}\cdots\right)^{m_{1}}. \nn \\
\eea
\normalsize To proceed, one should evaluate $R_1 ^{Empty} (\cdot), R_{2} ^{Empty} (\cdot),...$, as outlined in \ref{NumericalSimSub} below.

\subsection{Evaluation of  $R_{n}^{Empty}(\cdot)$}
\label{NumericalSimSub}

Consider once more the system illustrated in panel (a) of Fig. \ref{NumericsIllustration}. The probabilities $R_{1}^{Empty}(\cdot), R_{2}^{Empty}(\cdot),...$ are computed iteratively by simulating the dynamics of the walker in this system. 

We begin by computing $R_{1}^{Empty}(m)$ for $0\leq m \leq z-1$. We recall that $R_{1}^{Empty}(m)$ denotes the probability that exactly $m$ children nodes are accessible to the walker \emph{after} it arrives at the node in the first generation. For a child node in the second generation to be inaccessible, the node, as well as its children nodes, must all be occupied. Since this happens with probability $\rho ^z$, the probability for a \emph{specific} child node to be accessible is given by $1-\rho^{z}$. Note that the occupancies of the $z-1$ children nodes are independent due to the node in the first generation being empty by construction. Hence, the events of them being accessible to the walker are independent as well. The probability $R_{1}^{Empty}(m)$ is thus taken from the binomial distribution
\bea 
\label{R1}
R_{1}^{Empty}(m)=\binom{z-1}{m}(1-\rho^{z})^{m}\left ( \rho ^z \right )^{z-1-m},
\eea 
for $m = 0,...,z-1$. 

We now turn to the computation of $R_{2}^{Empty}(m)$, namely, the probability that exactly $m$ children nodes, of a specific node in the second generation, are accessible to the walker \emph{after} it has reached that specific node. Unlike the node in the first generation, which in our system was set to be unoccupied by construction, the $z-1$ nodes in the second generation are initially occupied with an obstacle density of $\rho$. Thus, when reaching a node in the second generation, the walker may push an obstacle to an unoccupied node in the third generation, increasing the obstacle density in the third generation and correlating between the occupancies of nodes there. A similar problem arises when computing $R_{n}^{Empty}(m)$ for higher generations. Thus, the binomial distribution in \eref{R1} no longer holds.  

To address this issue, we simulate the walker's dynamics for a large number of randomly generated systems, all from the type illustrated in panel (a) of Fig. \ref{NumericsIllustration}. For each system generated, we allow the walker to take a legal step, if possible, into each of the $z-1$ nodes in the second generation. Subsequently, for each node the walker reaches, we sample the obstacle occupancy of its children nodes in the third generation. This sample is represented by a binary vector of size $z-1$ entries, where $1$ indicates an occupied node and $0$ indicates an unoccupied node. Utilizing these samples, one can calculate the joint occupancy distribution of nodes in the third generation.

Having the above joint occupancy distribution in hand, we can compute $R_{2}^{Empty}(m)$. We do so by generating a new set of systems where the joint occupancy of nodes in the second generation is drawn from the distribution measured for third generation nodes in the previous iteration. The nodes in the third generation of the new systems are occupied with an obstacle density $\rho$. Once again, we allow the walker to take a legal step, if possible, into each of the $z-1$ nodes in the second generation. We then count the number of legal steps taken for each realization of the system. Using these statistics, we estimate the probability that a reached node in the second generation has $m$ accessible children nodes, namely, $R_{2} ^{Empty} (m)$. 

Repeating this process iterativley, one can estimate $R_{n}^{Empty}(\cdot)$ for $n=3, 4,..., g-1$. These probabilities can be substituted into \eref{suma} to estimate $Q_1 ^{Empty}$, from which $P_{Empty}$ follows immediately. In this work, estimates were made with $g = 400$, and $10^4$ system realizations for each iteration. As can be seen in Fig. \ref{ProbTheoryFig} and \ref{ScalingLawFig}, estimates are in excellent agreement with the analytical results. 

\subsection{Estimating $P_{Full}$}
To estimate $P_{Full}$ we follow the footsteps of the calculation of $P_{Empty}$, as outlined in \ref{PemptyCalc}, with minor differences described below. 

Consider a branch of a Bethe lattice with coordination number $z$. Nodes on the lattice are occupied with probability $\rho$, except for the node in generation $g=0$, which we leave empty, and the node in generation $g=1$, which is always occupied by an obstacle. We place the walker in generation $g=0$, as illustrated in panel (b) of Fig. \ref{NumericsIllustration}. Note that the only difference between this system and the system considered in \ref{PemptyCalc} is the occupancy of the node in generation $g=1$.

Similarly to the calculation of $P_{Empty}$, we calculate $P_{Full}$ using a sequence of auxiliary probabilities for the system in panel (b) of Fig. \ref{NumericsIllustration}: $E_1 ^{Full}, E_2 ^{Full}, E_3 ^{Full},...$. Here, $E_n^{Full}$, is the probability that a walker which reached a specific node in the $n$-th generation escapes to infinity through it, given it never returns to generation $n-1$. Note, that since the node in the first generation is now occupied, $E_n ^{Empty}$ and $E_n ^{Full}$ are generally different. For convenience, we denote the complimentary probabilities  $Q_n ^{Full} = 1 - E_n ^{Full}$. Note that $1 - P_{Full}=\rho ^{z-1}+\left (1-\rho ^{z-1}\right) Q_1 ^{Full} $, as we need to account for two scenarios: i) the node in the first generation is inaccessible; ii) the node in the first generation is accessible, but the walker does not escape to infinity through it. Thus, by estimating $Q_1 ^{Full}$, one also gets $P_{Full}$.

Following similar lines to those outlined in subsection \ref{PemptyCalc}, one can derive an analogous estimate for $Q_1 ^{Full}$. This can be done starting with  \eref{Q(1)}, replacing $Empty$ with $Full$ in the superscripts. Following this calculation, one ends up with an estimate for $Q_1 ^{Full}$ which is analogous to \eref{suma}. This estimate is independent of $Q_g ^{Full}$ and is given solely in terms of $R_{1}^{Full}(\cdot),...,R_{g-1}^{Full}(\cdot)$, which can be computed using the same algorithm described in \ref{NumericalSimSub}. Specifically, to compute the probability $R_{1}^{Full}(m)$, we start the \textit{Sokoban} walk at the origin and only consider instances in which it was able to get to the first generation. Using this sample, we estimate the probability that exactly $0\leq m \leq z-1$ children nodes are accessible to the walker \emph{after} it arrives at the node in the first generation.

\bibliography{ref}
\bibliographystyle{iopart-num}

\end{document}